\documentclass[prl,twocolumn,showpacs,amsmath,amssymb,superscriptaddress]{revtex4-2}

\usepackage{amsfonts} 
\usepackage{amsmath,amssymb,amsfonts,bm,color,graphicx,tabularx,physics}
\usepackage{balance}
\usepackage[unicode=true,colorlinks=true]{hyperref}
\usepackage[utf8]{inputenc}
\usepackage{textgreek}
\hypersetup{linkcolor=blue,citecolor=blue,urlcolor=blue}
\usepackage{diagbox}
\usepackage{color}

\newcommand{\beq}{\begin{equation}}
\newcommand{\eeq}{\end{equation}}
\newcommand{\beql}{\begin{equation*}}
\newcommand{\eeql}{\end{equation*}}
\newcommand{\beqn}{\begin{eqnarray}}
\newcommand{\eeqn}{\end{eqnarray}}

\renewcommand{\vec}[1]{\mbox{\boldmath$#1$}}

\begin{document}
\title{Robust Majorana Platform Driven by a Meissner-Induced Anisotropic Doppler Shift}
\author{Xiao-Hong Pan}
\affiliation{Tsung-Dao Lee Institute and School of Physics and Astronomy, Shanghai Jiao Tong University, Shanghai 201210, China}
\affiliation{College of Physics $\&$ Optoelectronic Engineering, Department of Physics, Jinan University, Guangzhou 510632, China.}

\author{Si-Qi Yu}
\affiliation{School of Physics and Institute for Quantum Science and Engineering, Huazhong University of Science and Technology, Wuhan, Hubei 430074, China}

\author{Li Chen}
\affiliation{Quantum Science Center of Guangdong-Hong Kong-Macao Greater Bay Area, Shenzhen 518045, China}

\author{Fu-Chun Zhang}
\email{fuchun@ucas.ac.cn}
\affiliation{Kavli Institute for Theoretical Sciences,  University of Chinese Academy of Sciences, Beijing 100190, China}

\author{Xin Liu}
\email{phyliuxin@hust.edu.cn}
\affiliation{Tsung-Dao Lee Institute and School of Physics and Astronomy, Shanghai Jiao Tong University, Shanghai 201210, China}
\affiliation{School of Physics and Institute for Quantum Science and Engineering, Huazhong University of Science and Technology, Wuhan, Hubei 430074, China}
\affiliation{Hefei National Laboratory, Hefei 230088, China}

\begin{abstract}
The realization of robust Majorana zero modes (MZMs), a cornerstone for fault-tolerant quantum computing, is hindered by the challenge of creating a platform that simultaneously offers a large topological gap, high tunability, and resilience to disorder. A system unifying these properties has remained elusive. Here, we propose and validate a novel platform that harnesses the Meissner effect in a topological insulator (TI) nanowire partially covered by a superconducting (SC) layer. Under an external magnetic field, Meissner screening currents in the SC induce a spatially varying Doppler shift on the TI surface. This effect generates a highly anisotropic effective g-factor, which selectively drives a topological phase transition localized on the nanowire's bottom surface. This mechanism is crucial as it spatially separates the topological phase from the SC/TI interface, permitting strong proximity-induced superconductivity while preventing detrimental band renormalization at the interface from closing the topological gap. Furthermore, by confining the topological superconducting phase to the gate-tunable bottom surface, our platform fully leverages the intrinsic disorder resilience of the TI's topologically protected surface states. Through a combination of supercurrent simulations, self-consistent Schrödinger-Poisson calculations, and large-scale tight-binding computations, we validate the platform's robustness. Our work establishes a practical pathway toward Meissner-mediated topological superconductivity for realizing robust MZMs in SC/TI hybrid systems.
\end{abstract}
\date{\today}
\maketitle

\section{I. introduction}

\begin{figure*}
    \centering
    \includegraphics[width=2\columnwidth]{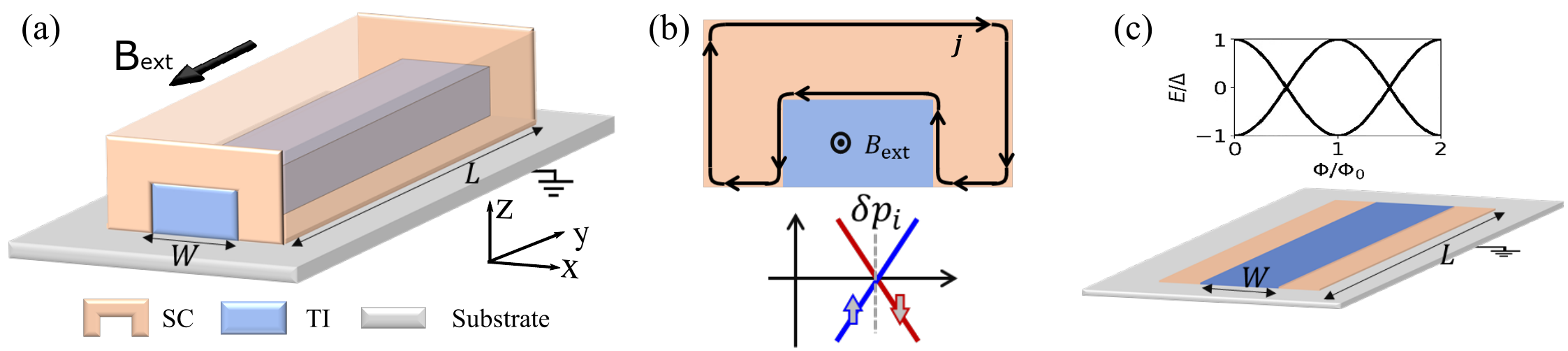}
    \caption{{\bf Meissner-Effect-Driven Majorana Platform.} (a) Schematic of the hybrid device, consisting of a topological insulator (TI) nanowire (blue) on a back gate (gray), partially covered by a conventional superconductor (SC, orange). (b) Device cross-section under an external magnetic field $\vec{B}_{\rm ext}$, forming a SQUID-like geometry. Screening supercurrents, 
\vec{j} (black arrows), generate a boundary-dependent Doppler shift. The inset schematically illustrates the resulting momentum shifts, $\delta p_i$, on the TI surface bands.
(c) View of the bottom TI surface, which hosts the low-energy physics. The top panel shows a representative Andreev spectrum at $k_y=0$ under periodic boundary conditions, indicating a topological phase transition.}
    \label{Setup}
\end{figure*}

Majorana zero modes (MZMs), exotic quasiparticle excitations that are their own antiparticles and exhibit non-Abelian braiding statistics, stand as cornerstone components for the development of fault-tolerant topological quantum computing~\cite{Beenakker2013,Sarma2015}. Since intrinsic topological superconductors are rare in nature, substantial theoretical and experimental efforts have focused on hybrid platforms that combine conventional $s$-wave superconductors (SC) with spin-orbit-coupled materials. These engineered systems, including semiconductor nanowires and topological insulators (TI), have yielded encouraging evidence of MZM~\cite{Fu2008,Fu2009,Sau2010,Lutchyn2010,Oreg2010,Cook2011,Deng2012,Rokhinson2012,Das2012,Wang2012,Churchill2013,Chang2015,Sun2016,Albrecht2016,Wiedenmann2016,Jeon2017,Liu2017,Zhang2017,Zhu2018,Liu2018, Volpez2019,Pan2019,Liu2020,Chen2021,Zhang2021,Papaj2021,Xie2021,Song2022,Li2022,Oshima2022,Chen2023}. However, a substantial gap remains between ideal theoretical models and experimental realizations, largely due to the complexity introduced by hybridization. Key challenges include: (i) the detrimental effect of disorder, which can create unintended states below the superconducting gap, complicating the identification of true MZMs; (ii) suppression of the proximity-induced superconducting gap by time-reversal symmetry-breaking element; and (iii) parameter renormalization at the interface, which can reduce the energy separation of excited states and even degrade the topological protection of the platform. Furthermore, the ability to manipulate and scale MZMs into complex networks remains limited. In this context, Majorana platforms based on nanowire geometries offer a promising path forward, as their layout flexibility aligns well with established schemes for Majorana manipulation and braiding.

In this work, we propose and theoretically validate a Meissner-driven Majorana platform designed to overcome these fundamental limitations. Our system comprises a three-dimensional (3D) TI nanowire on an electrostatic gate, partially covered by a conventional s-wave superconductor, as depicted in Fig.~\ref{Setup}(a). When an external magnetic field is applied, the Meissner effect induces screening currents in the superconductor. These currents, in turn, generate a boundary-dependent Doppler shift that functions as an effective Zeeman field~\cite{Reinthaler2015,Zhu2021,Pan2024,Liu2024}. The central feature of our proposal is that this field is engineered to be strong on the bottom surface of the TI nanowire but weak at the SC/TI interface (Fig.~\ref{Setup}(b)). This spatial separation drives a topological superconducting (TSC) phase transition predominantly at the bottom TI surface (Fig.~\ref{Setup}(c)), resolving a central conflict in hybrid devices: it permits strong superconducting proximity coupling at the SC/TI interface while independently inducing a robust topological phase at the gate-tunable boundary. Consequently, interface-induced band renormalization, typically a detrimental effect, is repurposed to enhance, rather than degrade, the topological protection gap. This architecture also intrinsically shields the topological state from interface roughness. The electrostatic gate provides direct control over the chemical potential at the bottom surface, leveraging the inherent robustness of TI surface states to enhance the resilience of the resulting MZMs against disorder and fluctuations. Our analysis is substantiated by a combination of realistic supercurrent simulations, self-consistent Schrödinger-Poisson calculations, and large-scale SC/TI tight-binding computations. Crucially, our proposed architecture aligns directly with recent experimental breakthroughs in fabricating bulk-insulating TI nanowires with partial superconducting coverage and integrated gates~\cite{Roessler2023}, establishing a clear and immediate path for experimental implementation.

The remainder of this paper is organized as follows. Section II details how the Meissner effect is engineered to generate the anisotropic Doppler shift and how the interface band renormalization enhances the stability of the proximity gap. Section III outlines the topological phase transition through analyzing the topoloigical phase transition under the Meissner effect. Section IV presents the main numerical results, first demonstrating the Meissner-driven topological phase transition and then the emergence and robustness of MZMs. Finally, Section V concludes by summarizing the findings and the application of this novel platform for future explorations of TSC and MZMs.

\section{II. The Meissner Effect and the Anisotropic Doppler Shift}

\label{sec:theory}
We consider a system comprising a TI nanowire partially covered by a conventional $s$-wave superconductor, situated on an electrostatic gate, as shown in Fig.~\ref{Setup}(a). An external magnetic field $B_{\rm ext}$ is applied parallel to the nanowire axis, which inherently induces the Meissner effect. Although this phenomenon is a fundamental property of superconductivity, its significance in hybrid structures is often underestimated. Recent experiments have revealed a significant Doppler shift in TI surface states, driven by the Meissner effect in fields as low as tens of mT \cite{Zhu2021}. This finding underscores the necessity of properly accounting for this phenomenon in any realistic model of SC/TI hybrids. In this section, we establish the theoretical framework for how the Meissner effect generates a highly anisotropic Doppler shift, which forms the basis of our proposal.

\subsection{II.A. Magnetic Field and Diamagnetic Current}
To determine the spatial distribution of the magnetic field and the diamagnetic screening currents, we self-consistently solve the London and Maxwell equations for our device geometry~\cite{supp, Bishop-VanHorn2022}. For a simply connected superconductor in a weak magnetic field, it is convenient to work in the London gauge ($\nabla \cdot \mathbf{A} = 0$), where the phase of the superconducting order parameter remains spatially uniform due to London's phase rigidity~\cite{Tinkham2004, Schrieffer1964}. The total magnetic field $\vec{B}$ is related to the diamagnetic current density $\vec{j}$ by the second London equation:
\begin{equation}\label{London}
\vec{B} = \vec{B}_{\rm ext} + \vec{B}_{\rm ind} = -\mu_0 \lambda_L^2 \nabla \times \vec{j},
\end{equation}
where $\vec{B}_{\rm ext}$ is the uniform external field, $\vec{B}_{\rm ind}$ is the induced field from the screening currents, $\mu_0$ is the vacuum permeability, and $\lambda_{L}$ is the London penetration depth. The induced field and the current are related by the Biot-Savart law:
\begin{equation}\label{B-S}
\vec{B}_{\rm ind}(\vec{r}) = \frac{\mu_0}{4\pi} \int_{V} \frac{\vec{j}(\vec{r}') \times (\vec{r}-\vec{r}')}{|\vec{r}-\vec{r}'|^3} dV'.
\end{equation}
Assuming translational invariance along the nanowire axis, we combine Eqs.~\eqref{London} and \eqref{B-S} to numerically solve for the magnetic field $\vec{B}$ and the current density $\vec{j}$ in the device cross-section~\cite{supp}. The results, shown in Fig.~\ref{Meissner}(a), confirm that the SC screens the magnetic field from its interior, with strong diamagnetic currents localized near the SC boundaries. Moreover, the current density profile in Fig.~\ref{Meissner}(b) (gray curve) confirms that the screening currents flow in opposite directions on the two sides of the superconductor, a behavior consistent with the analytical profile for a simple slab geometry (red curve)~\cite{Tinkham2004}.

In the London gauge, the current density is directly proportional to the magnetic vector potential, $\vec{j} = -\frac{1}{\mu_{0}\lambda_{L}^{2}} \vec{A}$~\cite{Tinkham2004}. After computing $\vec{A}$ within the SC, we extend the solution into the TI region by solving $\nabla\times\vec{A}=\vec{B}$ under the condition $\nabla\cdot\vec{A}=0$ with the appropriate boundary conditions~\cite{supp}. This vector potential couples to the charge carriers via the Peierls substitution, $\vec{p}\rightarrow \vec{p}-e \vec{A}$, inducing a momentum shift $\delta \vec{p} = e \vec{A}$ \cite{Zhu2021}. As shown in Fig.~\ref{Meissner}(c), this momentum shift is nearly uniform across the SC-contacted surfaces but is significantly larger on the TI bottom surface. This engineered spatial variation results in a strongly anisotropic Doppler shift on the TI surface states, a key feature of our platform.

\subsection{II.B. Bogoliubov-de Gennes Hamiltonian and Dressed Doppler Shift}
The Meissner effect is incorporated into the Bogoliubov-de Gennes (BdG) Hamiltonian of the hybrid system through the Peierls substitution:
\begin{equation}
H = \begin{pmatrix} H_{\rm TI}(\vec{A}) - \mu_{\rm TI}\tau_z & H_c^\dagger \\ H_c & H_{\rm SC}(\vec{A}) - \mu_{\rm SC}\tau_z \end{pmatrix},
\label{eq:BdG_Full}
\end{equation}
where $\tau_i$ are Pauli matrices in Nambu space. $H_{\rm TI}$ and $H_{\rm SC}$ are the BdG Hamiltonians in tight binding models for 3D TI (with a bulk electronic band gap $M$) and SC of the $s$-wave (with a superconducting pairing gap $\Delta$), respectively ~\cite{supp,Zhang2009,Liu2010}, with their kinetic terms modified by $\vec{A}$. The subsystems are coupled via a tunneling Hamiltonian $H_c$ with strength $t_c$, and $\mu_{\rm TI, SC}$ are the respective chemical potentials. Given the small magnetic fields considered, the direct Zeeman effect is negligible compared to the Meissner-induced effects and is omitted.

\begin{figure}[t]
    \centering
    \includegraphics[width=1\columnwidth]{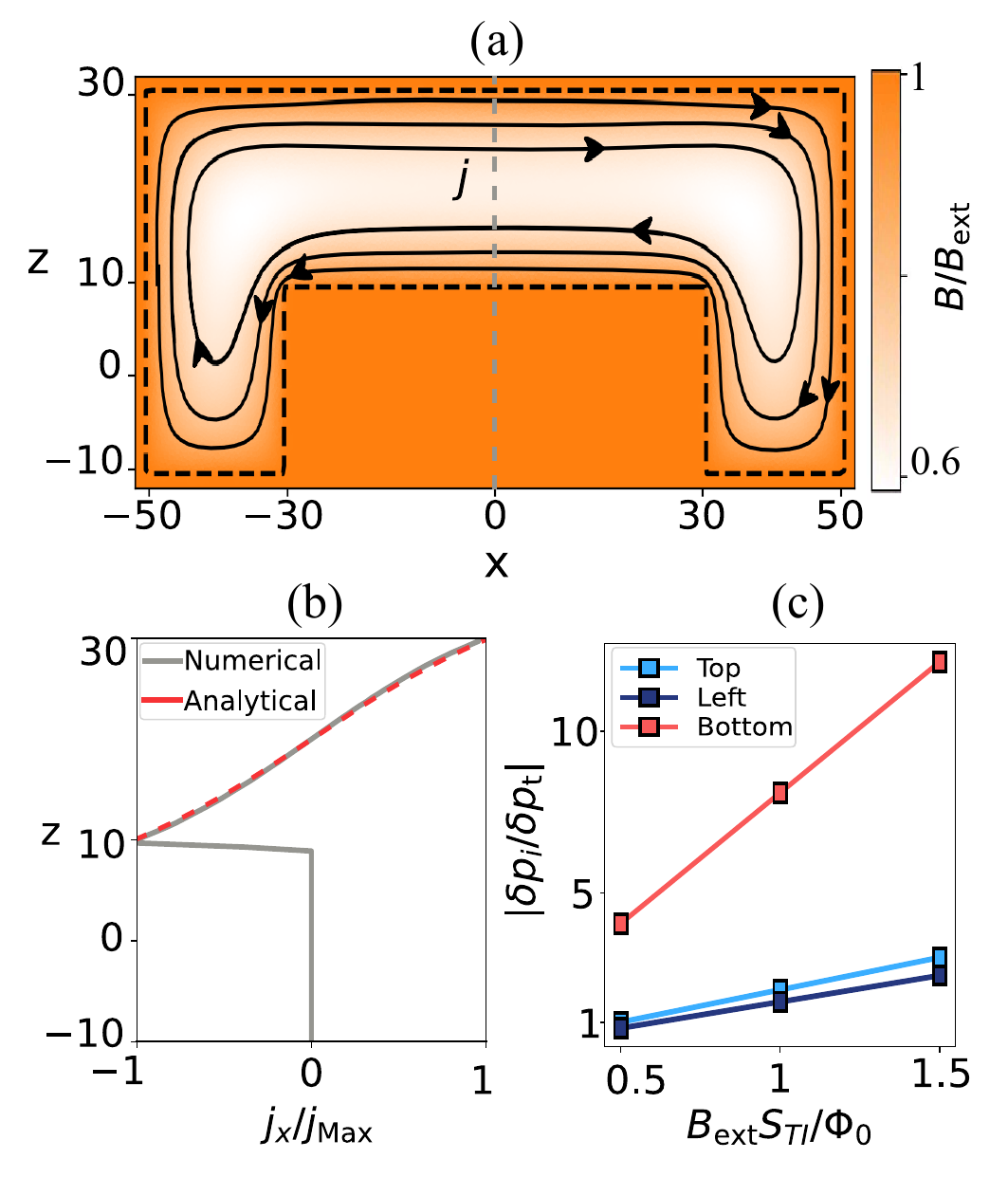}
    \caption{{\bf Meissner Effect and Anisotropic Momentum Shift.} 
    (a) Numerically calculated magnetic field $B$ (color plot, normalized to $B_{\rm ext}$) and diamagnetic current density $\vec{j}$ (arrows) in the device cross-section. The SC expels the magnetic field, generating strong screening currents near its boundaries. 
    (b) Distribution of the screening current density along the light gray path indicated in (a). For comparison, the red dotted line shows the analytical distribution for a superconductor with a simple slab geometry~\cite{Tinkham2004}.  (c) The resulting momentum shift $\delta p_i$ on the top, side, and bottom TI surfaces as a function of $B_{\rm ext}$. The shift on the bottom surface is significantly larger than on the SC-contacted surfaces, demonstrating the engineered anisotropy. Parameters are given in Supplemental Material~\cite{supp}.}
    \label{Meissner}
\end{figure}

\begin{figure*}[t]
    \centering
    \includegraphics[width=2\columnwidth]{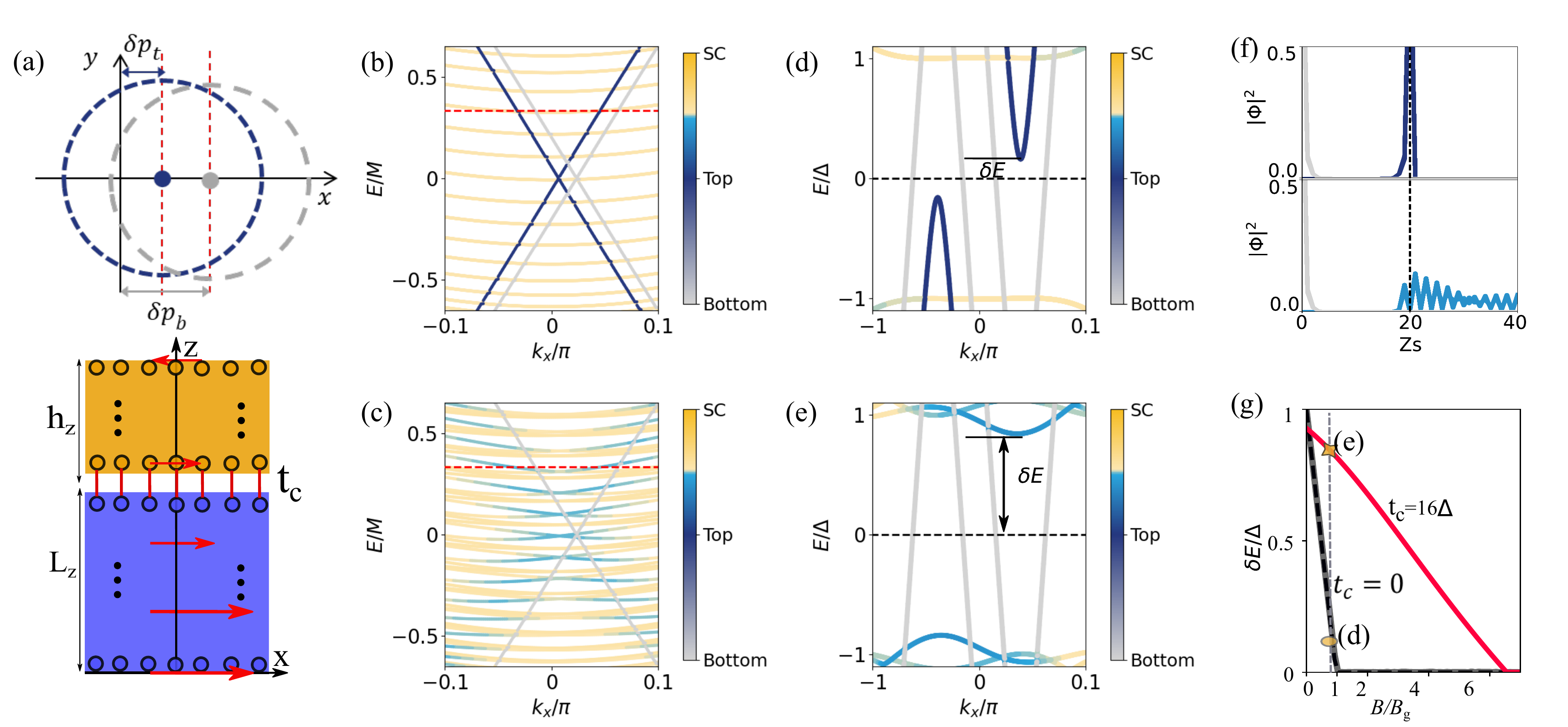}
    \caption{{\bf Proximity Gap Enhancement via Dressed Doppler Shift.}
    (a) Schematic illustration of
the Fermi surface of the TI (top panel) and the simplified SC/TI slab (bottom panel) used for analysis. The
interfacial coupling is denoted by $t_c$, and the red arrows indicate the vector potential distribution. The color of the bands in (b-f) indicates the average spatial position $\langle z \rangle$ of the eigenstates.
    (b) Uncoupled ($t_c=0$) band structure at finite $B_{\rm ext}$. The Dirac cone on the bottom surface (blue) shows a much larger momentum shift than the one at the top interface (gray).
    (c) Coupled ($t_c=16\Delta$) band structure. The top surface states hybridize strongly with the SC, forming Rashba-like bands (green), while the bottom surface states remain unaffected.
    (d) BdG spectrum for a reference model with weak coupling ($t_c=0$) and manually added pairing $\Delta$ on the top TI surface. The proximity gap is nearly closed by the Doppler shift.
    (e) BdG spectrum for the fully coupled system ($t_c=16\Delta$). The proximity gap at the interface remains large despite the same $B_{\rm ext}$.
    (f) Spatial distribution of the lowest-energy quasiparticle states for the weak-coupling (top) and strong-coupling (bottom) cases. In the strong-coupling case, the states are predominantly in the SC. (g) Proximity gap $\delta E$ at the top interface versus magnetic field. Strong coupling (red curve) significantly enhances the gap's resilience to the magnetic field compared to the weak-coupling reference model (black curve). The gray dashed line shows the expected gap closure from the analytical results in the weak coupling limit. Throughout we use the parameters for TI region mainly from Ref.~\cite{Zhang2009}. All parameters are detailed in the Supplemental Material~\cite{supp}.}
    \label{Dia_Cur}
\end{figure*}

{\it Band Structure and Dressed Doppler Shift-}
To analyze the essential physics of the Doppler shift, we simplify the device to the slab geometry shown in Fig.~\ref{Dia_Cur}(a). This simplification is justified because the momentum shift is nearly uniform across all SC-contacted surfaces (Fig.~\ref{Meissner}(b)). In this slab geometry, the vector potential $\vec{A}$ points along the $x$-direction and its magnitude increases with distance $z$ from the SC/TI interface~\cite{Tinkham2004,Reinthaler2015,Zhu2021}. In the weak-coupling limit ($t_c=0$), the low-energy physics is described by two decoupled Dirac cones for the top and bottom surfaces:
\begin{equation}\label{Doppler-1}
H_{\rm surf}(z) = v(z) [p_x - eA_x(z)] \sigma_y - v(z) p_y \sigma_x,
\end{equation}
where $\sigma_i$ are Pauli matrices in the spin basis and $v$ is the Dirac velocity. Fig.~\ref{Dia_Cur}(b) shows the band structure calculated from the full tight-binding model with $t_c=0$. As predicted by Eq.~\eqref{Doppler-1}, the momentum shift of the bottom surface's Dirac cone (blue) is much larger than that of the top surface (gray).

Upon introducing strong interfacial coupling ($t_c = 16\Delta$), the band structure is significantly renormalized, as seen in Fig.~\ref{Dia_Cur}(c). The states at the SC/TI interface no longer exhibit a Dirac dispersion; instead, they form Rashba-like bands (green) due to strong hybridization with the SC electrons. The effective spin-orbit coupling of these new bands is much smaller than that of the pristine TI surface states. We term the momentum shift experienced by these hybridized interface states as the "dressed Doppler shift"~\cite{Qiao2024}. This phenomenon, where hybridization alters material parameters like spin-orbit coupling, effective g-factor, etc., is also well-known in SC-semiconductor systems~\cite{Antipov2018,Woods2018,Vaitiekifmmodeeelseefinas2018,Mikkelsen2018,Winkler2019}. Crucially, the bottom TI surface remains far from the interface, and its electronic properties are unaffected by the coupling, thus preserving the large, bare Doppler shift. The combination of these effects yields a highly anisotropic response of the TI surface states to the magnetic field.

{\it Enhancement of the Proximity Gap-}
We now demonstrate how this engineered anisotropy enhances the robustness of the proximity-induced superconductivity. To quantify this, we compare the proximity gap in the weak- and strong-coupling regimes. As a reference for the weak-coupling limit ($t_c=0$), we manually insert a pairing potential $\Delta$ into the top third of the TI region. We apply a magnetic field large enough to nearly close this proximity gap via the Doppler shift, as shown in the BdG spectrum in Fig.~\ref{Dia_Cur}(d). In contrast, for the strongly coupled system ($t_c=16\Delta$, with pairing only inside the SC), the proximity gap at the interface remains wide open under the same magnetic field (Fig.~\ref{Dia_Cur}(e)). The quasiparticle states at the gap edge are now located predominantly within the SC (Fig.~\ref{Dia_Cur}(f)), indicating that the SC effectively "shields" the interface states from the momentum shift. Fig.~\ref{Dia_Cur}(g) confirms this conclusion by plotting the interface quasiparticle gap $\delta E$ as a function of the external magnetic field. Strong hybridization (black curve) leads to a much greater resilience of the proximity gap against the field compared to the weakly coupled reference model (red curve). Strong coupling mitigates the gap-closing effect of the Doppler shift.

{\it Effective anisotropic g-factor-}
This behavior can be uniformly described by defining the effective $g$-factor. The effective $g$-factor can be explicitly obtained in the weak-coupling limit by transforming the surface Dirac Hamiltonian of Eq.~\eqref{Doppler-1} to
\begin{equation}\label{Doppler-2}
H_{\rm surf} = v(z)(\hat{p}_x \sigma_y- \hat{p}_y \sigma_x) - \frac{1}{2}g^*\mu_b B_{\rm ext}\sigma_y,
\end{equation}
with $g^*=2evA_x(z)/\mu_b B_{\rm ext}$. The spatial dependence of $A_x(z)$ directly translates into an anisotropic $g^*$, which is weaker at SC/TI interface than at the bottom surface of the TI indicated by the slope of the curves in Fig.~\ref{Meissner}(b). The strong coupling between SC and TI further significantly suppresses the magnitude of $g^*$ at the SC/TI interface and enhances the superconducting proximity gap. As a result, magnetism (time-reversal symmetry breaking) and superconductivity are two key ingredients in implementing unpaired MZMs but coexist in a mutually competing relationship, with one increasing as the other decreases. In conventional platforms, these two ingredients are applied at the same location (the SC/normal-metal interface) so that a careful balance between these two factors must be struck to achieve TSC transitions at the interface. Our Meissner-driven platform avoids this dilemma by spatially separating the key ingredients through strong anisotropy g-factor. We use strong coupling at the SC/TI interface to induce a large, robust proximity gap. Simultaneously, we use the large, bare Doppler shift (i.e., a large effective $g^*$) at the bottom surface—far from the interface—to drive the topological phase transition. This architecture allows us to achieve a robust topological phase with a large energy gap under relatively weak external fields.

\section{III. Topological Phase Transition}
This section details the mechanism of the topological phase transition (TPT) in our platform. We begin with a low-energy effective model to provide an intuitive physical picture, followed by comprehensive numerical simulations that validate our claims in both idealized and realistic device environments.

\subsection{III.A. The Low-Energy Effective Model}
The engineered anisotropy of the effective $g$-factor is the central feature of our platform. It allows us to operate in a magnetic field regime where the topological phase transition is triggered on the bottom surface of the TI, while the SC/TI interface remains in a robustly gapped superconducting state. This spatial separation allows us to simplify the system to SC/TI-bottom/SC junction (Fig.~\ref{Setup}(c)). This effective system respects particle-hole symmetry and belongs to symmetry class D, whose topological nature is described by a $\mathbb{Z}_2$ invariant \cite{Kitaev2001}. With periodic boundary conditions (PBC), a change in the $\mathbb{Z}_2$ invariant is signaled by the closing and reopening of the quasiparticle gap at high-symmetry momenta $k_y=0$ or $k_y=\pi$ \cite{Kitaev2001}. Since the TI band inversion occurs at the $\Gamma$ point, the TPT in our system is expected at $k_y=0$. With the chemical potential inside the TI's bulk gap, the low-energy physics is governed by the helical edge states forming the 1D boundary of the 2D cross section (Fig.~\ref{Setup}(b)). The effective Hamiltonian for this 1D loop is:
\begin{equation}\label{Ham_edge}
H_{\rm eff} = v_{\Sigma}(-i\hbar\partial_{\Sigma}+eA_{\Sigma}\tau_{z}) \tau_{z}\tilde{s}_z-\mu\tau_{z}+\Delta_{\rm prox}\tau_x,
\end{equation}
where $\tilde{s}$ and $\tau$ are Pauli matrices for the edge-state and Nambu spaces, respectively; $v_\Sigma$ is the Fermi velocity along the loop direction $\Sigma$; and $A_\Sigma$ is the corresponding vector potential component. The proximity-induced pairing, $\Delta_{\rm prox}$, is finite only near the SC/TI contacts, creating a Josephson junction on the bottom surface of the TI.

Under a magnetic field, the Doppler shift $eA_\Sigma$ imparts a finite center-of-mass momentum to the Cooper pairs. As a pair traverses the closed loop along the edge of the TI cross section (Fig.~\ref{Setup}(b)), it acquires a gauge-invariant Aharonov-Bohm phase $\phi_{s} = \oint 2e\vec{A} \cdot d\vec{l}/\hbar = 2\pi \Phi/\Phi_0$, where $\Phi$ is the magnetic flux enclosed by the loop and $\Phi_0 = h/2e$ is the superconducting flux quantum. Due to the strong anisotropy of the Doppler shift, the TI bottom surface contributes the majority of this phase which behaves as a fractional Josephson junction with a $4\pi$-periodic energy-phase relation (Fig~\ref{Setup}(c)), $E \propto \pm\cos(\phi_{s}/2)$~\cite{Fu2009a}. The TPT occurs when the gap closes, which happens precisely at $\phi_s = (2n+1)\pi$, corresponding to a half-integer flux $\Phi = (n+1/2)\Phi_0$ \cite{Kitaev2001,Fu2009}. Crucially, this condition for the TPT depends only on the enclosed flux, making it fundamentally independent of the chemical potential.

\subsection{III.B. Numerical Validation in an Idealized System}

\begin{figure*}[t]
    \centering
    \includegraphics[width=2\columnwidth]{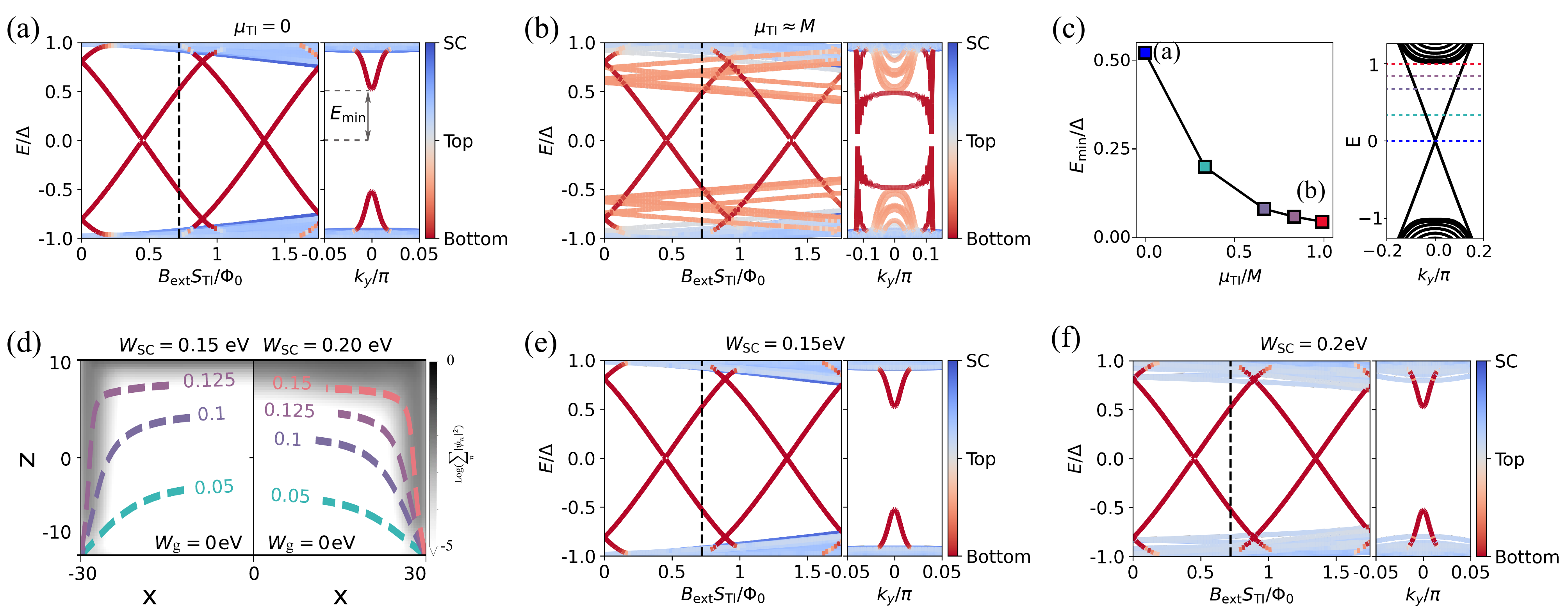}
    \caption{{\bf Topological Phase Transition and Robustness to Chemical Potential.}
    \textbf{(a-c) Uniform Chemical Potential:} Analysis based on the full tight-binding model assuming a spatially uniform chemical potential $\mu_{\rm TI}$.
    (a), (b) Left: Low-energy spectrum at $k_y=0$ as a function of magnetic flux $\Phi$ for $\mu_{\rm TI}=0$ (Dirac point) and $\mu_{\rm TI}=0.99M$ (near bulk band edge), respectively. The red curves, localized on the bottom surface, show a clear gap closing and reopening, signaling a TPT. The blue curves, localized at the SC/TI interface, remain gapped. Right: Full BdG band structure at the flux indicated by the vertical dashed line in the left panel.
    (c) Minimum quasiparticle gap as a function of $\mu_{\rm TI}$ at a fixed flux within the topological phase ($\Phi=0.8\Phi_0$). The gap is largest near the Dirac point. The right panel shows the TI band structure for reference.
    \textbf{(d-f) Non-Uniform Potential:} Analysis using a realistic, self-consistently calculated electrostatic potential $U(x,z)$ with the gate voltage fixed at $W_g=0$.
    (d) Calculated electrostatic potential profile (contours) and the spatial distribution of states at the Fermi level for two different SC/TI work function offsets, $W_{\rm SC}$.
    (e), (f) Left: Low-energy spectrum versus flux for $W_{\rm SC}=0.15$ eV and $W_{\rm SC}=0.2$ eV. Right: Corresponding BdG band structures. Despite the complex potential profile, the TPT on the bottom surface remains robust, and the topological gap is large. The color of all curves indicates the average spatial position of the eigenstate. Parameters are detailed in Supplemental Material~\cite{supp}.}
    \label{Topo_phase}
\end{figure*}

To validate the predictions of our low-energy model, we now present numerical simulations using the full tight-binding (TB) Hamiltonian from Eq.~\eqref{eq:BdG_Full} for the device geometry in Fig.~\ref{Setup}(a) with PBC along the $y$-direction. We first consider an idealized case with a spatially uniform chemical potential $\mu_{\rm TI}$.

Fig.~\ref{Topo_phase}(a) shows the result for $\mu_{\rm TI}=0$ (at the Dirac point). The left panel plots the low-energy spectrum at $k_y=0$ versus the applied magnetic flux. The states localized on the bottom surface (red curves) exhibit a distinct gap closing and reopening, a hallmark of a TPT. In contrast, the states at the SC/TI interface (blue curves) remain fully gapped, protected by the strong proximity effect and the suppressed Doppler shift. The right panel confirms that the minimum gap occurs at $k_y=0$, consistent with our analysis.

Next, we test the robustness of the TPT against changes in the chemical potential. Fig.~\ref{Topo_phase}(b) shows the same calculation for $\mu_{\rm TI}=0.99M$, placing the Fermi level near the bulk conduction band edge. As predicted by the flux-driven mechanism, the TPT occurs at nearly the same magnetic field and remains confined to the bottom surface. However, the overall size of the topological gap is dramatically reduced around $k_f$ which is similar with the gap drop in the topological planar Josephson junction \cite{Pientka2017}. As shown in Fig.~\ref{Topo_phase}(c), the topological gap is maximal when $\mu_{\rm TI}$ is at the Dirac point and decreases as the Fermi level moves towards the bulk bands. This result highlights that while the TPT itself is robust, achieving the largest possible topological gap requires tuning the chemical potential on the bottom surface towards the Dirac point—a task for which our gate-controlled architecture is ideally suited.

\subsection{III.C. Robustness in a Realistic Electrostatic Environment}

In a real device, the chemical potential is not uniform. A work function mismatch between the SC and TI materials creates a band offset, $W_{\rm SC}$, and the bottom gate voltage, $W_g$, imposes an external electrostatic potential. Notably, the difference of the work function between SC and TI, and can be raised above the
electronic TI band gap \cite{Xu2014}. The presence of the inhomogeneous electrostatic potential can significantly affect the band structure and electron distribution at the SC/TI interface so as to form TSC \cite{Pan2024,Antipov2018,Mikkelsen2018,Woods2018,Chen2024}.   To model this, we self-consistently solve the coupled Schrödinger-Poisson equations to obtain a realistic, non-uniform electrostatic potential profile $U(x,z)$ within the TI region \cite{Antipov2018, Mikkelsen2018, Woods2018}. The details of this calculation, which treats the SC and gate as fixed boundary conditions, are provided in Supplemental Material~\cite{supp}.

\begin{figure*}[t]
    \centering
    \includegraphics[width=1.8\columnwidth]{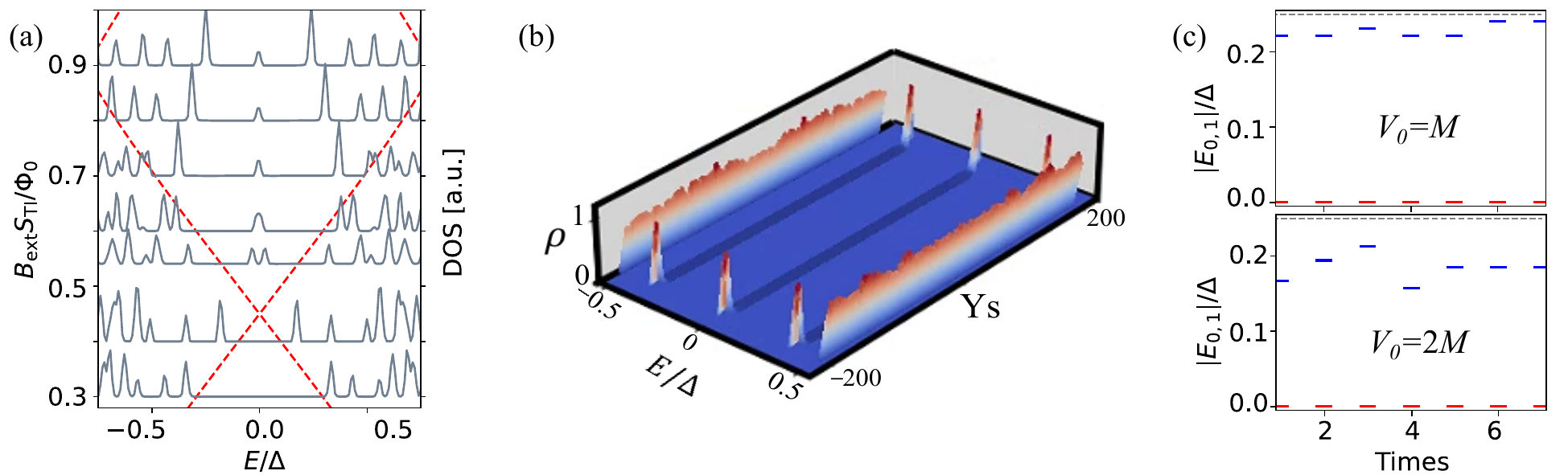}
    \caption{{\bf Emergence and Robustness of Majorana Zero Modes.}
    (a) Density of states (DOS) at the end of the nanowire as a function of energy and magnetic flux. A sharp zero-energy peak emerges and stabilizes after the topological phase transition at $\Phi = 0.6\Phi_0$.
    (b) Spatial distribution (probability density) of the three lowest-energy states at $\Phi=0.7\Phi_0$. The zero-energy state (middle) is clearly resolved into two MZMs localized at opposite ends of the nanowire. The first excited states and bulk states are delocalized.
    (c) Stability of the three lowest-energy modes against disorder. The plot shows the energies of the three DOS peaks closest to $E=0$ across several random impurity configurations for $V_0=M$ (top) and $V_0=2M$ (bottom) disorder strengths. The first excited energy remains finite, demonstrating robust topological protection. The parameters are identical to those in Fig. \ref{Topo_phase}.}
    \label{MZM}
\end{figure*}

We set the gate voltage to $W_g=0$ to align the chemical potential at the bottom surface with the Dirac point, which is optimal for a large topological gap. We then consider two values for the band offset at the SC/TI interface: $W_{\rm SC}=0.15~\text{eV} = M$ and $W_{\rm SC}=0.2~\text{eV} > M$. The resulting potential profiles and Fermi-level state distributions are shown in Fig.~\ref{Topo_phase}(d). The potential decreases from $U \approx W_{\rm SC}$ at the top interface to $U \approx W_g$ at the bottom. This spatially varying potential confines the electronic states, preventing them from extending far from the interfaces, even when the local chemical potential enters the bulk bands.

We incorporate this self-consistent potential $U(x,z)$ into our TB Hamiltonian and re-calculate the spectrum. The results are shown in Figs.~\ref{Topo_phase}(e) and \ref{Topo_phase}(f). Remarkably, the system's behavior remains robust. The proximity gap at the SC/TI interface is large, enhanced by the potential-induced confinement of states near the SC. The TPT still occurs on the bottom surface, and because the gate holds the local chemical potential at the Dirac point, the resulting topological gap is nearly identical to the optimal gap found in the idealized $\mu_{\rm TI}=0$ case. Even when $W_{\rm SC}$ is large enough to populate bulk bands near the interface (right panel of Fig.~\ref{Topo_phase}(d)), the topological properties remain stable. These results confirm that our platform's core principle—the spatial separation of a robust proximity gap and a gate-tunable TPT—is not an artifact of an idealized model but persists in a realistic electrostatic environment.

\section{IV. Emergence and Stability of Majorana Zero Modes}
\label{sec:mzm}
Having established the existence of a robust topological phase, we now investigate the properties of MZMs under open boundary condition in all directions. We computed the density of states (DOS) as a function of energies under varied magnetic field strength through the Python package TBPLaS~\cite{Li2023}.

Fig.~\ref{MZM}(a) plots the DOS spectrum as a function of applied magnetic flux. For fluxes below the critical value $\Phi_c \approx \Phi_0/2$, the system is fully gapped, with no states inside the bulk topological gap (demarcated by red dashed lines, extracted from Fig.~\ref{Topo_phase}(f)). With magnetic flux larger than $\Phi_0/2$, the DOS peak at $E=0$ appears, which is consistent with the TPT analysis in Sec.~III. The minor splitting of this peak near the transition point, $\Phi=0.55 \Phi_0$, is a finite-size effect, which rapidly diminishes as the system moves deeper into the topological phase, forming a stable zero-energy mode.

Within the topological regime, a series of in-gap excited states also appear, whose energies depend non-monotonically on the magnetic field. This behavior contrasts with the evolution of the bulk gap highlighted by the red dashed line, suggesting that these in-gap states are Andreev bound states (ABSs) accompanied by the emergence of the MZMs. The characteristics of the lowest ABSs are crucial because its energy dictates the magnitude of the topologically protected band gap. To confirm the identity and location of the low-energy modes, we computed their spatial distributions. Fig.~\ref{MZM}(b) shows the probability densities for the three lowest-energy states at a representative flux $\Phi=0.7\Phi_0$. The zero-energy ground state is clearly composed of two MZMs, each tightly localized at an opposite end of the nanowire. The first excited states are also end-localized, consistent with their interpretation as the lowest-energy ABSs of the SC/TI nanowire.

A defining characteristic of MZMs is their topological protection against local perturbations. This is a critical property, as experimental TI nanowires are known to contain charged impurities that can cause significant chemical potential fluctuations~\cite{Roessler2023,Lippertz2024,Schluck2024,Nikodem2024,Nikodem2024a,Brede2024}. While our analysis in Sec.~III indicates that the TPT is insensitive to the smooth chemical potential distribution, robustness against random spatial fluctuations must be explicitly tested. To this end, we introduce on-site potential disorder by adding a random term $V_{\rm dis}(\vec{r})$ to the TI Hamiltonian, drawn from a uniform distribution $V_{\rm dis}(\vec{r})\in[-V_{0}/2,V_{0}/2]$. We test disorder strengths up to $V_0=2M$, which significantly exceeds the expected experimental fluctuation scales up to tens of meV \cite{Brede2024}.

Fig.~\ref{MZM}(c) demonstrates the remarkable outcome of this test. It plots the absolute values of the three DOS peaks closest to $E=0$ in several different configurations of random disorder. Even under strong disorder ($V_0 = 2M$), the zero energy peak remains robust while the first excited state peak remains finite. This profound stability provides direct, compelling evidence that our platform hosts genuinely protected MZMs. This robustness is a direct consequence of our design: the TPT is governed by a global quantity (the magnetic flux), and the topological states are formed from the intrinsically protected TI surface states, which are shielded from interface-specific effects and gate-tuned to an optimal operating point.

\section{V. CONCLUSION}
\label{sec:conclusion}
In summary, we have proposed and theoretically validated a novel platform to achieve robust Majorana zero modes in gated SC/TI hybrid nanowires. We have shown that the Meissner effect, often considered a complication, can be harnessed to create a highly effective stabilization mechanism. By generating a strongly anisotropic Doppler shift, our architecture spatially separates the region of strong superconducting proximity coupling (the SC/TI interface) from the region where the topological phase transition occurs (the TI's bottom surface).
This design paradigm resolves a central conflict in the engineering of topological superconductors, allowing a large, hard proximity gap to coexist with the magnetism required for the topological phase. Through large-scale numerical simulations that incorporate realistic electrostatic effects, we have demonstrated that the resulting topological phase and its emergent MZMs are exceptionally robust. They are resilient to uniform chemical potential variation, largely insensitive to the non-uniform potentials from band bending, and, most importantly, are protected against strong on-site disorder.
Given the recent, rapid experimental progress in fabricating precisely controlled SC/TI hybrid devices of the type we have modeled~\cite{Bai2022,Roessler2023,Schluck2024,Nikodem2024,Nikodem2024a}, our work establishes a practical and promising route toward developing the stable, scalable, and topologically protected qubits required for fault-tolerant quantum computing.


%

\end{document}